\documentclass[usegraphicx]{mn2e}

\title{{\it XMM-Newton} observations of UW CrB -- detection of X-ray bursts and evidence for accretion disc evolution}
\author[]{Pasi Hakala$^{1}$\thanks{E-mail:
pahakala@astro.helsinki.fi},
Gavin Ramsay$^{2}$,
Panu Muhli$^{1}$,
Phil Charles$^{3}$,
\newauthor
Diana Hannikainen$^{1}$,
Koji Mukai$^{4}$,
Osmi Vilhu$^{1}$ \\
\\
$^{1}$Observatory, P.O. BOX 14, University of Helsinki, FIN-00014 University of Helsinki, Finland.\\
$^{2}$Mullard Space Science Laboratory, University College London, Holmbury St Mary, 
Dorking, Surrey, RH5 6NT, UK.\\
$^{3}$School of Physics \& Astronomy, University of Southampton, Southampton SO17 1BJ, UK.\\
$^{4}$Laboratory for High Energy Astrophysics, NASA/GSFC, Code 662, Greenbelt, MD 20771, USA\\}
\begin{document}

\date{}

\pagerange{\pageref{firstpage}--\pageref{lastpage}} \pubyear{2004}

\maketitle

\label{firstpage}

\begin{abstract} 

UW CrB (MS1603+2600) is a peculiar short period X-ray binary that
exhibits extraordinary optical behaviour. The optical light curve
shape of the system changes drastically from night to night, without
any changes in overall brightness.  Here we report X-ray observations of UW CrB 
obtained with {\it XMM-Newton}. We find evidence for several X-ray bursts confirming a neutron
star primary. This considerably strengthens the case that UW CrB is 
an Accretion Disc Corona (ADC) system located  at a distance of at least 5--7 kpc,
 (3--5 kpc above the galactic plane). The X-ray and optical monitor
(UV+optical) light curves show remarkable shape variation from one
observing run to another, which we suggest are due to large scale
variations in the accretion disc shape resulting from a warp which
periodically obscures the optical and soft X-ray emission. This is also supported
by the changes in phase-resolved X-ray spectra.

\end{abstract}


\section{Introduction}

UW CrB (previously known as MS1603+2600) was discovered by Morris
et al. (1990) in the \textit{Einstein} medium sensitivity
X-ray survey.  They identified the optical counterpart with a blue,
emission-line only object close to the X-ray position. Subsequent
optical photometry revealed a period of 111 minutes. Morris et
al. (1990) also noted UW CrB's key distinguishing feature of an
optical light curve that changes dramatically from one night to
another.  UW CrB is located at a high galactic latitude of 47$^o$,
which, together with its low X-ray flux, suggests that either
the source is underluminous in X-rays for a low mass X-ray binary
(LMXB) or that it is actually located in the galactic halo.

\begin{figure*}
\includegraphics[scale=0.9]{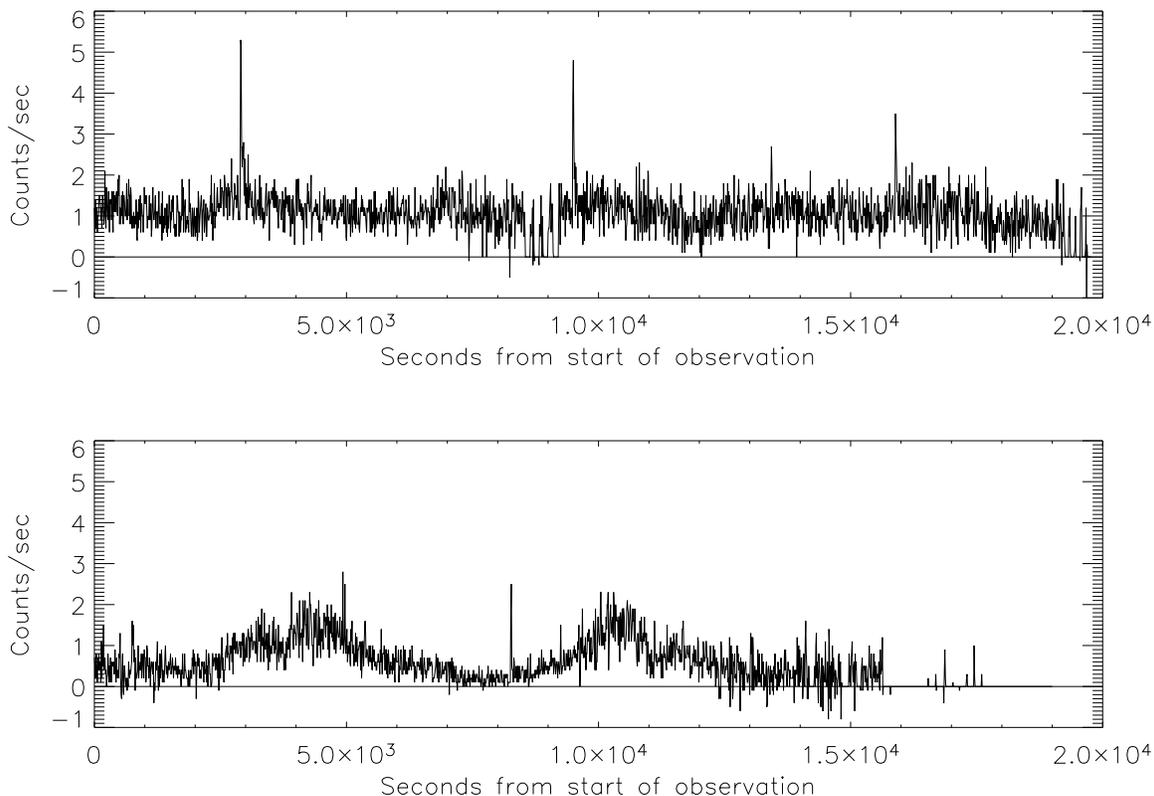}
 \caption{The background subtracted unfolded EPIC pn 0.2--10 keV light curves from epoch 1 (top, 2003 Jan 20) and epoch 2 (bottom, 2003, Jan 22). The time resolution is 10 sec. Several X-ray bursts are seen during each observation. }
\end{figure*}

\begin{figure*}
\includegraphics[scale=0.9]{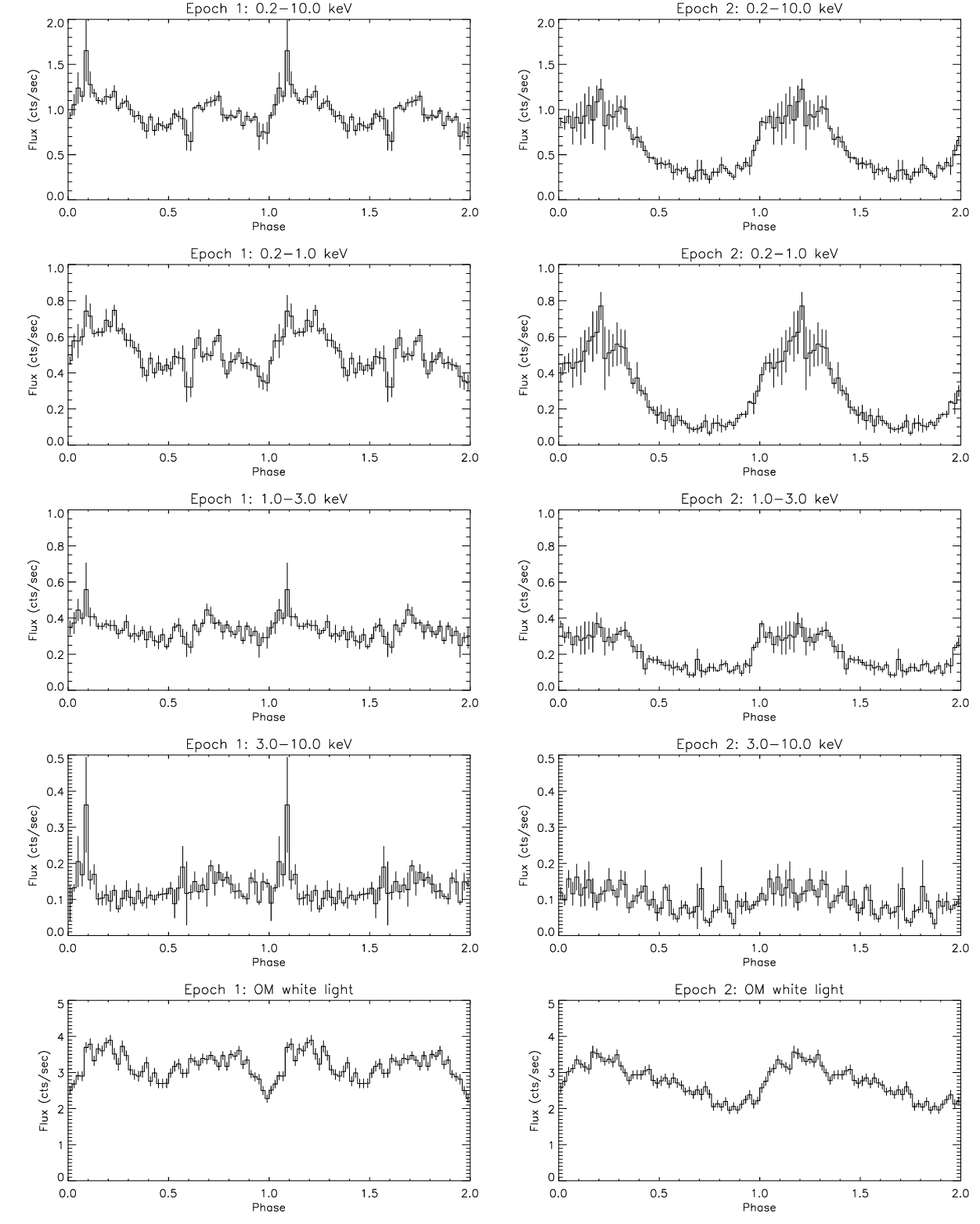}
 \caption{The phase folded EPIC-pn and OM lightcurves. From top to bottom: 0.2--10.0 keV, 0.2--1.0 keV,
1.0--3.0 keV, 3.0--10.0 keV and OM white light. The left panel is epoch 1 and the right panel epoch 2.
The 0.0 phase is arbitrary and taken from the main sharp minimum of epoch 1 OM light curve. }
\end{figure*}

Hakala et al. (1998) published the results of further optical
photometry together with the ROSAT observations.  They confirmed the
extreme variability seen in the optical light curve pulse shapes. Even
if there are dramatic changes in the optical light curve shape, the
overall optical flux level does {\bf not} vary significantly.  ROSAT data showed
that the soft X-ray spectrum of UW CrB can be modelled with a single
blackbody component, with a temperature of kT$\sim$0.24 keV. As
expected at this galactic latitude, the spectra showed no significant
interstellar absorption.

Earlier models for UW CrB have invoked both LMXB and magnetic CV
(cataclysmic variable) explanations (Morris et al. 1990, Ergma \& Vilhu 1993). 
 Hakala et al.  (1998) compared the ROSAT spectra of
UW CrB with soft X-ray spectra of various types of interacting
binaries. Their conclusion was that the only class where the spectral
fits seemed to match those of the source was the (black hole) soft
X-ray transients in quiescence. This could also have explained the
relatively low \textit{F$_{X}$/F$_{opt}$} ratio ($\sim$15)
observed.  In addition, Hakala et al.  (1998) report a negative
result from their search for circular polarisation.

Mukai et al. (2001) presented ASCA observations of UW
CrB. They claim to see a single type I X-ray burst, which would rule
out both black hole and white dwarf explanations for the accreting
component in the system.  They proposed instead that UW CrB is a short
period X-ray dipper, similar to X1916-05. They also note that the
system would probably in this case have to reside in the galactic
halo, and could have formed in the globular cluster Palomar 14, which
is located 11$^o$ from UW CrB and at a distance of 73.8 kpc.

Recently Jonker et al. (2003) report the results from Chandra
observations of the source. They conclude that depending on the nature
of the single burst event detected by ASCA (Mukai et al. 2001), UW CrB
could be either a ``nearby'' quiescent transient or an ADC source at 11--24 kpc
distance. They consider the dipper interpretation of Mukai
et al. (2001) to be unlikely, mainly on the basis of optical
brightness (eg. the system would be intrinsically much brighter in the
optical than other, longer period LMXB's that have a larger disc).
Quite recently, Muhli et al. (2004) reported that the optical light curves of
UW CrB show optical bursts. This has now also been confirmed by
Hynes et al. (2004).  

In this paper we will present new results of UW CrB taken using {\it XMM-Newton}
and discuss the nature of the source in light of these extensive new data.

\section{Observations}

UW CrB was observed by {\it XMM-Newton} on 2003 Jan 20 and 22. The two
datasets  produced a total integration time of 38.8 ksec. However, the particle background 
was high for around half of observation.  
The observations were processed using SAS v5.4.1. and the light curves and
background subtracted spectra were extracted using XMMSELECT. 
Response files prepared by the {\it XMM-Newton} Project Team were
downloaded and used in the EPIC pn spectral analysis, which was carried
out using XSPEC. We only used events with flag=0 and pattern=0--4 in the
spectral analysis, whilst events with pattern=0--12 were included in
the light curves. The spectra were binned using GRPPHA with a minimum
requirement of 30 counts per spectral bin. The light curves were also 
background subtracted.

\section{Light curves and bursts}
\subsection{X-ray and UV-optical light curves}

The unfolded X-ray lightcurves are shown for the two epochs in
Figure 1.  There are two main features in the light curves that
characterize the variability. Firstly, at least 5--6 short, type
I-like, X-ray bursts are seen during both of the epochs.
Furthermore, it is clear from the unfolded time series that
the light curves have changed within the 2 days that separate the two
observing runs. Whilst the light curve of the first observation shows only
a moderate modulation over the 111 minute orbital period,
such a modulation is prominent in the second dataset, as is
more clearly demonstrated in the phase-folded light curves (Figure 2).
These show that the orbital modulation, while present in both
observations, has changed remarkably from one observation to
another. The modulation is most pronounced at lower energies and UV-optical
(OM white light $\sim$ 2000--7000\AA). The cause for the  orbital modulation 
will be discussed in more detail in the Discussion section.  


\subsection{X-ray bursts}

The candidate X-ray bursts seen in the raw X-ray time series are, at
least sometimes, also evident in the OM white light data. There are
at least 5--6 candidate bursts in our data.  However, the start time can 
be determined only for the strongest (first)  burst.   An analysis of that
burst revealed that, based on our 1 sec time resolution light curves, the burst 
seems to start about 2 sec later in UV-optical, than it does in the X-rays (Figure 3). 
Given the 111 minute orbital period, this roughly corresponds to the light
crossing time of the system. This burst happened at an approximate binary
phase of 0.1, which implies that the UV-optical burst  could occur as a result
of reprocessing in the parts of the disc "behind" the compact object. We must
note though that the binary phase here is not entirely certain, but
based on the deepest minimum in the OM light curve (see Figure 2). It
is possible that, given the extent of the changes in the light curve
shapes of UW CrB, the deepest minimum does not always occur
when the compact object is closest to be eclipsed.

\begin{figure}
\includegraphics[scale=0.5]{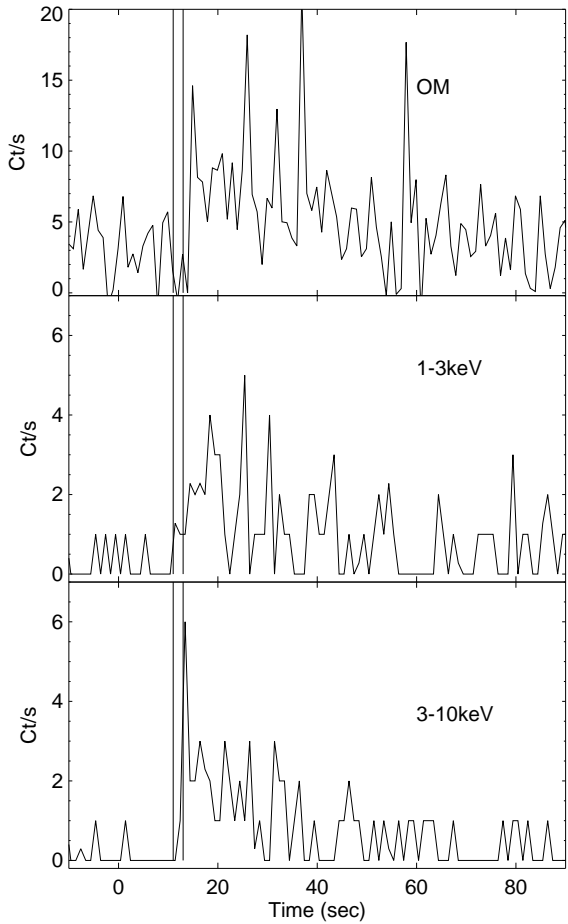}
\caption{The best (first) burst of the first run at different energies. 
There is no detection in 0.2-1 keV band.}
\end{figure}

The peak EPIC pn count rate of our best (first) X-ray burst is 6 cps, this together
with the powerlaw spectral model in the 1--10 keV range (photon index
1.74 from the summed spectrum for epoch 1) yields $2.44 \times
10^{-11} $ergs cm$^{-2}$  s$^{-1}$ using PIMMS. Now if we assume a
conservative burst peak flux of 10$^{37}$ ergs s$^{-1}$, this would
imply a distance of 60 kpc, and this would still be a lower limit, as
10$^{37}$ ergs s$^{-1}$ is on the low side of estimated type I burst
peak fluxes (Lewin et al 1995). In addition we are using the brightest
of our candidate bursts for this estimate. Based on burst fluxes at
different energies we conclude that their spectrum could be
characterized by a 1.5--2 keV blackbody, which is fairly typical for
type I bursts. The fact that bursts are not seen below 1 keV is a
natural consequence of their X-ray spectrum.

It is possible, that if UW CrB really is an ADC source, as suggested
by Jonker et al. (2003), then the
scattered flux we see from the burst could be only a fraction of the
intrinsic flux.  Assuming we detect only $\sim$1\% of the total
flux implies the derived minimum distance estimate would
reduce to about 6 kpc.  However, there is a problem with this
interpretation in that we are still seeing a sharp burst rise of
duration less than 1 sec and we would expect the scattering ADC to smooth this
out. Having said that, we think this does not entirely rule out the
ADC scattering scenario. UW CrB is a short period system, and an ADC
could have a diameter with light crossing time of less than 1 sec, in which
case the smearing of pulse could not be detected in our data.  There is some
additional support for the ADC model from the burst data in that the
bursts during the first run, when there is little orbital
modulation, have a peak flux of $\sim$ 6 cps, whilst the bursts during
the second run have a peak flux of only 2--3 cps. This suggests that we
are not seeing the bursts directly, but maybe only the scattered
X-rays from an ADC, which is more obscured during the second
epoch. This is also supported by the fact that the orbital X-ray
modulation is more pronounced during the second epoch.

\section{X-ray spectra}

\begin{table*}
 \caption{EPIC pn spectral fits to the two summed spectra and to the ``high'' and ``low'' phase spectra of epoch 2.
The corresponding $\chi_\nu^2$ values are 1.21, 1.22, 1.23, 1.14. The values shown for different parameters indicate 90\% confidence intervals.}
 \label{symbols}
 \begin{tabular}{@{}lccccccc}
  \hline
  Spectrum & $N_H$ $\times$ 10$^{22}$cm$^{-2}$ & kT$_{BB}$ (keV) & norm. (BB) & kT$_{mekal}$ (keV) & norm. (MEKAL) &$\alpha$ (pl) &  norm. (pl) \\
  & & & $\times 10^{-5}$ & & $\times 10^{-4}$ & & $\times 10^{-4}$ \\
  \hline
 
 Obs. 1. & 0.076-0.152 & 0.098-0.142 & 0.93-3.18 & 0.996-1.199 & 1.37-2.43 &    1.477-1.659 & 3.42-4.29  \\
 Obs. 2. & 0.083-0.197 & 0.089-0.125 &  0.78-3.88  &  1.030-1.215 & 1.38-2.35 & 1.190-1.433 & 1.50-2.03 \\
 Obs. 2. (high) & 0.075-0.181 & 0.084-0.118 & 1.09-5.75 & 0.944-1.098 & 1.93-3.30 &1.410-1.784 & 2.08-3.16 \\
 Obs. 2. (low) & 0.073-0.360 & 0.082-0.196 & 0.05-4.62 & 1.041-1.510 & 0.42-1.92 &0.944-1.435 & 0.81-1.52  \\ 
  \hline
 \end{tabular}
\end{table*}

The X-ray spectra of UW CrB have been previously interpreted using a
variety of models. Hakala et al. (1998) showed that the ROSAT spectra
could be modelled using either a 0.24 keV blackbody, or a 1.0 keV
thermal bremsstrahlung model. Jonker et al. (2003) fitted their
Chandra data using a powerlaw with photon index 2.

Our summed EPIC pn spectrum from the first observing run is plotted in
Figure 4, and the resulting spectral fit parameters, together with those for
the second observing run, are shown in Table 1. All spectra were extracted 
only from times of low background and have integration times of 6.1 and 6.9 ksec 
for the two epochs respectively.
 We find that in order to be able to fit the spectra, a rather
complex model is required. The whole 0.2--10.0 keV EPIC pn spectrum
{\it cannot} be fitted with a single blackbody model nor a powerlaw
model. Even the combination of these two models, previously used to
model the ROSAT and Chandra spectra, cannot produce an adequate
fit. As a result we show that an additional thermal plasma component is
required in order to fit the spectra.  

\begin{figure}
\includegraphics[scale=0.35, angle=-90]{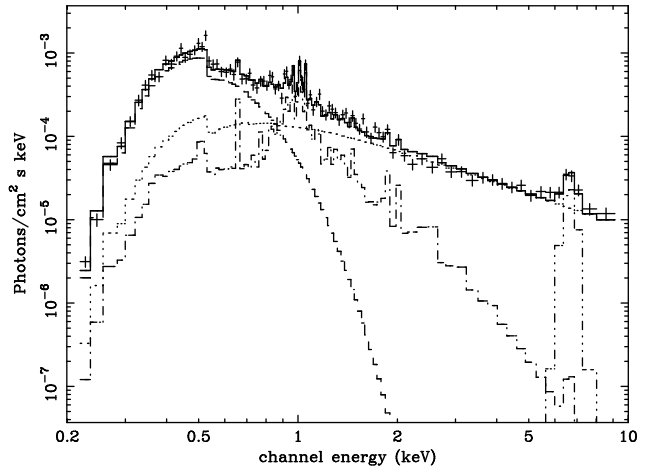}
\caption{The summed EPIC pn spectra of epoch 1. }
\end{figure}

We have also looked at the MOS1 spectra from the first observing
run. Fitting the MOS1 spectra we get somewhat different results than
from the pn spectra. Especially the derived spectral components that
dominate at the low energy range (photoelectric absorption and the
blackbody temperature) are different. The MOS1 fits yield blackbody
temperatures above 0.2 keV and $N_H$ values less than 0.05$\times$
10$^{22}$ cm$^{-2}$, whilst the pn spectrum (taken simultaneously) implies
 T$_{bb} <$0.15 keV and $N_H$ $> 0.07\times$ 10$^{22}$ cm$^{-2}$ respectively (99\%
limits). The remaining spectral components (MEKAL and powerlaw) seem
to match.  It is known that the EPIC MOS and EPIC pn count rates
differ by 10--15\% below 0.7 keV (Kirsch et al. 2004) . We think
that the differences in spectral fits can probably be explained by 
such systematic calibration effects. Furthermore, fitting first the hard part
(above 2 keV) of the MOS spectrum with just a powerlaw, fixing
its value, and then adding a blackbody component fixed to the 
best fit value from the pn spectrum and fitting the whole spectrum
yielded $\chi_{\nu}^{2}=$ 1.76. Adding a MEKAL component
and letting the powerlaw to change freely brought the $\chi_{\nu}^{2}$
value down to 1.11, which is a reasonable fit. With this fit (blackbody 
temperature still fixed to 0.1 keV) all the spectral parameters agree
with the pn data. 

\begin{figure}
\includegraphics[scale=0.35, angle=-90]{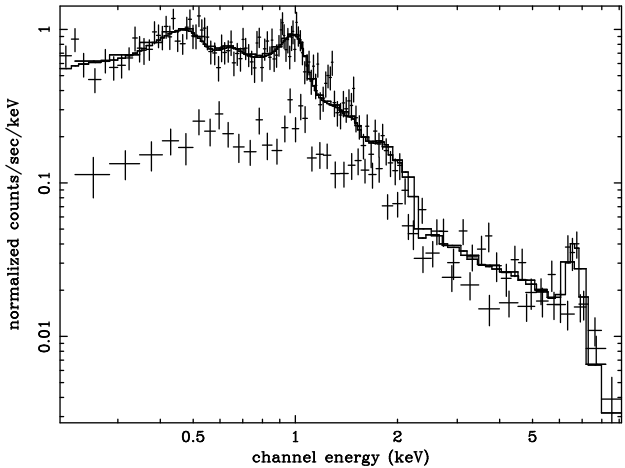}
\caption{The ``high'' and ``low'' phase  EPIC pn spectra from the second epoch (together
with a spectral model fit to the ``high" spectrum). }
\end{figure}

The spectral fits to the datasets from the two epochs  give very similar
results, in spite of the totally different light curves. This is
because most of the flux in both spectra is contributed to by
the ``high'' part of the light curves, and thus the effect of
phase-dependent absorption is diminished. There is a slight
difference in the powerlaw photon index though. However, the phase-resolved
spectra extracted separately during the low and high parts of the
light curve (phases 0.6-0.9 and 0.1-0.35 respectively) from the second epoch {\it do} show a clear
difference (Figure 5).  The fits to these spectra are also included
in Table 1. We fitted the higher signal to noise spectrum ("high" spectrum ) first, 
and then fixed all the other fitting parameters except for the column density and 
the normalizations for the fitting of the "low" spectrum.  We determined that the changes
over orbital phase can be well reproduced by simply changing the
overall flux of each component (i.e. through their normalisations).
While we had not anticipated this given the energy dependence of the
X-ray orbital modulation (which had suggested that the most likely
cause for the modulation would be through a changing $N_H$ with
orbital phase), a closer inspection of Figure 5 does show that the low
energy spectral shape remains roughly constant while decreasing by a
factor $\sim$5 in flux.



We suggest that this is most likely caused by changing partial covering effects across
the disc by very thick matter. The normalisations of all three
spectral components are affected by this, even though it is strongest
in soft X-rays.  An obvious cause for such an effect could be the
raised rim (or warp) in a non-axisymmetric accretion disc. This "rim
asymmetry", combined with an  extended X-ray emission source (ADC)
is the most likely cause for the orbital modulation. The fact that the 
powerlaw component appears steeper during the "high" spectrum 
suggests that the ADC is likely to be hotter in its outer regions than
near the disc surface. Such effect has been theoretically predicted by 
Ko \& Kallman (1994)

In summary, the main difference for the the ``high" and ``low" 
spectra is in the normalisation of the components. However, the
powerlaw component of the ``high"  spectrum is somewhat steeper than
in the ``low" spectrum indicative of a temperature gradient within the 
ADC.  This is also supported by the integrated
spectra from the two epochs. The powerlaw at epoch 1 (when there is
less orbital modulation, and thus less obscuration of the inner disc on
the average) is somewhat steeper than at epoch 2.   

We have also inspected the RGS spectra, but due to the low count rates
the spectra cannot be analyzed in detail. However, there is evidence for
emission lines at around 0.5keV and possibly also near 0.65 keV. 
These could correspond to NVII and OVIII Ly$_{\alpha}$ type lines. 
These  lines have also been seen for instance in the RGS spectra of
Her X-1 (Jimenez-Garate et al., 2002).  

\section{Discussion}

Perhaps the most intriguing result coming out of our study is the huge
variation in the amount and shape of the X-ray modulation between the
two epochs. It is well known from previous optical studies by
Morris et al. (1990) and Hakala et al. (1998) that the optical light
curve shape varies even from night to night. Our study here
demonstrates that similar variability is also seen in X-rays.

It is evident from both the X-ray modulation shape (and its
variability) and phase-resolved X-ray spectra, that the most likely
cause is the changing total column depth as the binary system rotates.
 This is most simply produced by non-axisymmetric vertical structure in the outer parts of
the accretion disc. Observational evidence for such structure has been
reported in many LMXBs like X1822-371 (Mason \& Cordova 1982, Hellier
\& Mason 1989), X1916-05 (Callanan 1993; Homer et al 2001), AC211
(Ilovaisky et al 1993) and X1957+115 (Hakala, Muhli \& Dubus,
1998). Similar structure has also been seen in supersoft sources such
as CAL 87 (Schandl et al. 1997).  Sufficiently thick disc rims are
almost impossible to produce theoretically, so an alternative
explanation for the X-ray (and optical) modulation is the warped disc
model, where a thin disc is warped out of the orbital plane due to
radiation pressure effects (Pringle, 1996). The precession of such a
warped disc could then manifest itself through the changing light
curve shape over a (longer) precession period. However, calculations by
Ogilvie \& Dubus (2001) show that radiation-induced warping is not very likely
to happen in short period LMXBs like UW CrB. Instead they suggest that
the likely cause for long term variability in short period systems is the
change in accretion rate through the disc. It is, however, hard to reconcile
how such an effect could produce the observed X-ray and optical light curves,
which clearly favour an explanation, where a non-axisymmetric, vertically extended
obscuring structure is required.  In fact our recent
optical work on UW CrB (Hakala et al. in prep.) suggests that there
could be another (about 5d) period in UW CrB, over which the optical
light curve shape seems to be repeating.  Now, our two sets of
{\it XMM-Newton} observations presented here were taken 2d apart, very close
to half of this suggested periodicity. This could explain the very
different X-ray light curve shapes reported here.

During the first epoch we see the dip or partial eclipse at
phase 0.0 predominantly in the soft band. This implies, together with
less X-ray orbital modulation, that we can see the inner disc (or
parts of it) directly. There is an additional dip at phase 0.5, which
seems to imply extra (obscuring) matter at 180$^o$ from the
secondary. Similar features have been seen in optical light curves of
X1916-05 (Callanan 1993). The OM light curves during this epoch
are very similar in shape to the soft X-ray light curves. This implies
that both are probably dominated by the emission from the inner
disc. The hard X-rays are not affected by the disc structure. During
the second epoch the soft X-ray light curve is more
sinusoidal with a minimum just before phase 0.0. This is what would be
produced by a thick disc rim that has a bulge on the side where the
stream hits the outer disc.  This time no eclipse is seen in soft
X-rays or UV-optical, which could be a result of the outer disc
obstructing our view of the inner disc. The fact that we see a large 
smooth X-ray modulation over the orbital period is a hallmark of
ADC systems.

\section{Conclusions}

Using our {\it XMM-Newton} data, we find evidence for evolving vertical disc structure (either warped or an 
asymmetric flared disc). This is evident from the strong X-ray modulation over the orbital period, that 
has changed amplitude and shape over the 2 day period separating the two epochs. The nature of 
X-ray and UV-optical modulation also suggests that UW CrB is most likely an ADC source.  The detection 
of several X-ray bursts confirms that the compact object in UW CrB is a neutron star. However, as we 
probably only see scattered X-rays from the ADC, the bursts luminosities cannot be used as a direct tool for
distance estimates. Assuming that we would only see about 1 \% of the total burst flux  then we can
estimate that the minimum distance to the source would be $\sim$ 6 kpc. This would place the source
at about 4 kpc above the galactic plane.

\section*{Acknowledgments}

 PJH, PM and DH are supported by the
Academy of Finland.

\end{document}